# Password Cracking and Countermeasures in Computer Security: A Survey

Lifeng Han

The University of Manchester, Oxford Raod, Manchester, UK

lifeng.han@manchester.ac.uk

*Abstract*—**With the rapid development of internet technologies, social networks, and other related areas, user authentication becomes more and more important to protect the data of the users. Password authentication is one of the widely used methods to achieve authentication for legal users and defense against intruders. There have been many password-cracking methods developed during the past years, and people have been designing countermeasures against password cracking all the time. However, we find that the survey work on password cracking research has not been done very much. This paper is mainly to give a brief review of the password cracking methods, import technologies of password cracking, and the countermeasures against password cracking that are usually designed at two stages including the password design stage (e.g. user education, dynamic password, use of tokens, computer generations) and after the design (e.g. reactive password checking, proactive password checking, password encryption, access control). The main objective of this work is to offer the abecedarian IT security professionals and the common audiences some knowledge about computer security and password cracking and promote the development of this area.**

## I. INTRODUCTION

The password has been used to encrypt information or message for a long time in history and it leads to a discipline: cryptography [1]. Furthermore, with the rapid development of computer science, password is now also commonly used for user authentication issues, which is very important to internet security [2] [3].

RFC 2828 defines user authentication as "the process of verifying an identity claimed by or for a system entity". The authentication service must assure that the connection is not interfered with by a third party masquerading as one of the two legitimate parties, which usually concerns two approaches data origin authentication and peer entity authentication [4]. The data origin authentication provides for the corroboration of the source of a data unit without the protection against the duplication or modification of data units, and this type of service supports applications like email where there are no prior interactions between the communicating entities. The peer entity authentication provides for the corroboration of the identity of a peer entity in an association for use of a connection at the establishment or at times during the data transfer phase, which attempts to provide confidence that an entity is not performing either a masquerade or an unauthorized replay of a previous connection.

There are usually four means of authenticating user identity based on: something the individual knows (e.g. password, PIN, answers to prearranged questions), something the individual possesses (token, e.g. smartcard, electronic keycard, physical key), something the individual is (static biometrics, e.g. fingerprint, retina, face), and something the individual does (dynamic biometrics, e.g. voice pattern, handwriting, typing rhythm) [5][6][4]. In the different methods, password authentication (something the individual knows) is a widely used line of defense against intruders [7]. In the password authentication scheme, the user ID determines that the user is authorized to access the system and the user's privileges. It is also sometimes used in discretionary access control meaning that others can log in using your identity. When the user provides a name/login and password the system compares the password with the one stored for that specified login. However, some users do not pay attention to the confidentiality or complexity of their passwords thinking that they do not have any private files on the internet. This allows the malicious crackers to make damage the entire system if they provide an entry point to the system [8]. Furthermore, higher-speed computational processors have made the threats of system crackers, data theft, and data corruption easier than before [9].

This paper is constructed as below: Section 2 introduces the related work about computer security and password cracking; Section 3 introduces the password application in remote user authentication and the vulnerabilities; Section 4 presents the password cracking methods including traditional methods and the import technologies from other literature such as mathematics and linguistics; Section 5 presents the countermeasures for password cracking and Section 6 shows the conclusions closely followed with the appendix of available software.

## II. Related Works

Morris and Thompson [10] conduct a brief case study of password security and specify that the salted passwords (12-bit random quantity) are less predictable at that time. Jobusch and Oldehoeft [11][12] show that password mechanisms remain the predominant method of identifying computer system users. They first give a review of the authentication, discuss the vulnerabilities of password mechanisms at that time, and then describe an extension of the UNIX password system with the permission of the use of pass-phrases. Spafford [13][14] designs a sufficiently complex password screening dictionary OPUS to prevent weak password choices with a space-efficient method. Cazier and Medlin [15] present research on password security and password cracking time focusing on e-commerce passwords. Their empirical investigation shows that e-commerce users usually create passwords without guidance from the website and they tend to create short or simple passwords such that they can easily remember them. Marechal [16] presents some existing hardware architectures of that time dedicated to password cracking, such as Field-programmable gate arrays (FPGAs) which are devices containing fully programmable logic and CELL processor (a core processor of the PlayStation3 console) that are developed by Sony, Toshiba, and IBM. Snyder [17] designs a pattern for creating individualized learning exercises in the area of security education with password cracking/recovery realized in practice. Furnell [18] discusses the website password protection issues in three aspects including whether the sites provide any guidance when users select passwords, whether any restrictions are imposed on passwords, and whether offers means to help users who have forgotten their passwords. Plaga [19] conducts research on the suitable use cases of biometric keys. Winder [20] briefly introduces some of the most used password-cracking methods including dictionary attacks, brute-force attacks, rainbow table attacks, malware, offline cracking, and guessing, etc.

Other related works include Bonneau [21] conducting an analysis of an anonymized corpus of 70 million passwords; Kelley et al. [22] measuring the password strength by simulating password-cracking algorithms; Zhang et al. [23] introducing a time-memory-resource trade-off method for password recovery; Klein [24] describing a collection of attacks used by surfers and web sites against other web sites, as well as attacks by web sites against surfers; Houshmand and Aggarwal [25] introducing the probabilistic techniques to enhance the passwords.

Different from the previous work, this paper will give discussions on both passwords cracking methods and countermeasures against password cracking. Furthermore, to read this paper, you do not need any background knowledge. Thus, we hope it can offer the abecedarian IT security professionals and the common audiences some knowledge about computer security and password cracking, and promote the development of this area.

## III. Password in Remote User Authentication

In remote user authentication, password protocol is used when the password is employed for authentication [2] [4]. In the password protocol scheme, the user first transmits identity to the remote host; then the host generates a random number (nonce) and the random number is returned to the user, at the same time host stores a hash code of the password; the user hashes the password with the random number (a random number helps defend against an adversary capturing the user's transmission) and sends it to the host; the host compares the stored information with the received one from the user to check whether they math or not. Remote user authentication is the authentication over a network, the internet, or a communications link. The additional security threats of remote user authentication include eavesdropping, capturing a password, and replaying an authentication sequence that has been observed. Remote user authentication generally relies on some form of a challenge-response protocol to counter threats.

## IV. Password Cracking Methods

### A. *Traditional Password Cracking Methods*

Since passwords remain the most widely used mechanism to authenticate users, obtaining passwords is still a common and effective attack approach [26]. The traditional password cracking methods include stealing, defrauding, user analysis, algorithm analysis, and fully guessing, etc. which will be introduced below.

#### *1) Stealing*

Password stealing can be achieved by looking around the person's desk, shoulder surfing, sniffing the connection to the network to acquire unencrypted passwords, gaining access to a password database and malware. In shoulder surfing, hackers will take the guise of a parcel courier, service technician, or something else to make them get access to an office. After they get in the building, the service personnel uniform provides a kind of free pass to wander around and make note of passwords being entered by genuine members of staff [20]. In the malware attack, the key logger or screen scraper is usually installed by malware that can record everything the user type and take screenshots during the login process. Furthermore, some malware tries to look for the existence of a web browser client password file and copy the access passwords from the browsing history.

#### *2) Defrauding*

Another way to gain the password is to defraud the users by social engineering or phishing online. Social engineering will take the jugglery in the real world. One case is to telephone an office pretending as a professional IT security technicist and ask for the users' account (or network) access passwords [15]. In the phishing attack, users will receive a phishing email that contains links leading to faked websites such as online banking and payment, etc., and makes some terrible problems to the accounts' security.

*3) User Analysis*

Users tend to generate passwords based on the things they usually chat about on social networks or somewhere else. Password crackers are likely to look at this kind of information and make a few guesses during the cracking of passwords. The hackers can reduce the password cracking time by the analysis of the special users according to their characteristics, such as the persons' names, job titles, interests, families, hobbies, and so on. One such kind of attack is known as Spidering. The hackers realize that many corporate passwords are generated by connecting to the business itself. They try to build custom word lists by studying corporate literature, website material, listed customers, etc.

TABLE I. AVERAGE TIME REQUIRED FOR EXHAUSTIVE KEY SEARCH [4] © STALLINGS AND BROWN L (2011)

| Key size (bits) | Number of alternative keys | Time required at 1 Decryption/µs | Time required at 106 Decryptions/µs |
|---|---|---|---|
| 32 | $2^{32} = 4.3\times10^{9}$ | $2^{31}\mu s = 35.8\ minutes$ | 2.15 milliseconds |
| 56 | $2^{56} = 7.2\times10^{16}$ | $2^{55}\mu s = 1142\ yeas$ | 10.01hours |
| 128 | $2^{128} = 3.4\times10^{38}$ | $2^{127}\mu s = 5.4\times10^{24}\ years$ | $5.4\times10^{18}\ years$ |
| 168 | $2^{168} = 3.7\times10^{50}$ | $2^{167}\mu s = 5.9\times10^{36}$ years | $5.9\times10^{30}$ years |
| 26 characters (permutation) | $26! = 4\times10^{26}$ | $2\times10^{26}\mu s = 6.4\times10^{12}$ years | $6.4\times10^{6}$ years |

TABLE II. COMPARISON OF THREE POPULAR SYMMETRIC ENCRYPTION ALGORITHMS [4], DES: DATA ENCRYPTION STANDARD; AES: ADVANCED ENCRYPTION STANDARD © STALLINGS AND BROWN L (2011)

| | DES | Triple DES | AES |
|---|---|---|---|
| Plaintext block size (bits) | 64 | 64 | 128 |
| Ciphertext block size (bits) | 64 | 64 | 128 |
| Key size (bits) | 56 | 112 or 168 | 128,192, or 256 |

*4) Algorithm Analysis*

Algorithm-analysis attacks focus on the used encryption algorithms such as the cryptanalytic attacks which are also used in the decryption of ciphertext [9]. It relies on the nature of the algorithm, some knowledge of the general characteristics of the text, and some sample of plaintext-ciphertext pairs. This kind of attack exploits the characteristics of the algorithms to attempt to deduce a specific plaintext or the keys.

*5) Fully Guesting*

The widely used kind of method for password cracking is fully guessing including the dictionary attack, brute-force attack, hybrid of dictionary and brute-force [15], rainbow table attacks, etc.

In the dictionary attack, a large dictionary containing possible passwords (commonly used passwords, e.g. the common dictionary words, the combination of several words the pun, etc.) is used by the hackers attempting to gain the access to the user's computer or network [27]. The common approach is applying the same encryption method to the dictionary of passwords to compare with the copy of an encrypted file containing the passwords. If the encryption method has used the hash function, then each dictionary password must be hashed using each salt value to compare with the stored hash values. Berger et al. [28] also introduce dictionary attacks using keyboard acoustic emanations.

A brute-force attack guesses the password using a random approach by trying different passwords and hoping one works [29]. This approach is different from the dictionary attack with the use of non-dictionary words, which can contain all possible alpha-numeric even special character combinations, such as "aaaaa000", "zzzzz9999" and "bbb&&99$00". If some logic analysis is employed in the brute-force attack the password cracking time may be reduced sometimes, e.g. the use of previously mentioned user analysis (using persons' names, job information, and hobbies). Assuming that the users' passwords are over a handful of characters long, the brute-force attack can crack the passwords eventually even though it costs a lot of time. The use of distributed computing models and zombie botnets can shorten the cracking time. A brute-force attack is also used in the decryption of ciphertext where it tries all possible keys on the ciphertext until an intelligible translation output of plaintext is gained. As supplementary, Table 1 and Table 2 show the example of the average time required for a brute-force attack on symmetric encryption and the key size of several popular symmetric encryption algorithms.

In the rainbow table attacks, the hackers use a rainbow table that is a list of pre-computed hash values for all encrypted passwords with all salts [30]. The time that rainbow table attack takes to crack a password is reduced to the time it takes to look it up in the list. This kind of attack could be countered by the use of a sufficiently large salt value and a sufficiently large hash length. Rainbow table attacks have been used widely, e.g. [31], [32], and MD5 online cracking using rainbow tables[1].

If password cracking is conducted online, the protection system usually locks out the users after they failed to log in to the system more than three times to block the automated guessing software. However, offline attacks usually escape from this. In an offline attack, hackers first break into a system to steal the encrypted password files or eavesdrop on an encrypted exchange on the internet. Then the password cracker can take as long as they need to try and crack the code without alerting the target system or individual user.

[1] http://www.passcracking.com/

### B. Import Technology of Other Literature

Some algorithms or models are imported from other literature to be combined with the existing password-cracking methodologies, such as the mathematical Markov model and the linguistic context-free grammar.

#### 6) Markov Chains Based Method

Markov models are used in some of the password crackers or recovery tools, such as [33], [34], the John the Ripper[2] password cracker [35], and AccessData's Password Recovery Tool [36]. The Markov chains can reduce the search space which will be needed if the brute-force attack is used.

Narayanan and Shmatikov [33] use the Markovian filters to capture the phonetic similarity with words in the users' native languages (only seeking the "memorable" strings ) and employ the finite automated model to deal with the non-alphabetic characters in the passwords. The used Markov models include the zero-order model and the first-order model. In the zero-order Markov model, characters are independent of each other and they are generated according to the probability distribution with the description formula (P is used for the Markovian probability, v as the frequency of the character):

$$P(\alpha) = \prod_{x \in \alpha} v(x)\; P(a) \; = \backslash pai \tag{1}$$

On the other hand, the characters in the first-order Markov model are dependent and each character will be assigned a probability by looking at the previous characters (symbol "|" representing that the right part is the premise of the left part):

$$P\left(x_1 x_2, ..., x_n\right) = v(x_1) \prod_{i=1}^{n-1} v(x_{i+1} | x_i) \tag{2}$$

In their experiments, they test the approaches on the data of 150 user passwords and recover 67.6% coverage using a search space of size $2\times10^9$ compared with the Oechslin's attack [31] that achieves 27.5%. Marechal [16] presents that Markov chains can also be used as a powerful tool for improving distributed and rainbow table cracking.

### C. Probabilistic Context-Free Grammars

Weir et al. [37] employ Probabilistic Context-Free Grammars (PCFG) for password cracking. They first create a probabilistic context-free grammar which is based on a training set of disclosed passwords. They incorporate the probability distribution information of user passwords to generate password patterns. They use grammar to generate word-mangling rules which will be employed in the password-guessing period with a dictionary attack. The material they used includes three word-lists that are MySpace test list (containing 33481 plaintext words), the SilentWhisper list (7480 plaintext passwords), and the Finnish test list (22733 unique MD5 password hashes) and six input dictionaries, of which four dictionaries (English_lower, Finnish_lower, Swedish_lower, and common_Passwords) are from John the Ripper with other two dictionaries "dic-0294" (from a password-cracking site[3]) and "English_Wiki" (user updated dictionaries[4]). Their method has shown improvement of 28% to 129% as compared to a publicly available password cracking program "John the Ripper" testing on a set of disclosed passwords. The running time information is that it can generate 1144895.1 unhashed guesses per second using a computer equipped with "MaxOSX 2.2GHz Intel Core 2 Duo". Rao et al. [38] also introduce a work about the effect of grammar on the security of long passwords.

## V. Countermeasures for Password Cracking

The protection of passwords from compromise and unauthorized use is a crucial issue since passwords remain the most popular approach for authentication. The countermeasures for password cracking could be achieved in two stages generally, i.e. the password design stage and after the generation.

### D. Password Design Stage

#### 7) User Education

Users can be educated on the importance of using strong passwords and be trained on how to generate hard-to-guess passwords using some password selection strategies. For instance, the password must contain letters (both capital Letters and small Letters), numbers, and special characters; the length of the password must not be less than a certain number; the use of a passphrase which is usually longer than a word; the use of beginning characters of each word in a memorable sentence, etc. Shay et al. [39] emphasize the importance of user attitudes and behaviors for stronger passwords. Komanduri et al. [40] conduct research to measure the effect of password-composition policies (e.g., requiring passwords to contain symbols and numbers), and explore the relationship between password-composition policies and the strength of the resulting passwords. They characterize the predictability of passwords by calculating the entropy and find that some commonly held beliefs about password composition and strength are inaccurate. They give a contribution by several recommendations for password-composition policies that result in strong passwords.

#### 8) Dynamic Password

[2] http://www.openwall.com/john/

[3] http://www.outpost9.com/files/WordLists.html

[4] www.wiktionary.org

Cole [26] gives the introduction of the one-time password, dynamic password, and static password, of which one-time password is one way to provide a high level of security. In a one-time password scheme, a new password is required each time when the user logs on to the account to prevent the hackers from using a pre-compromised password. The dynamic password specifies that the password is changed frequently or at a short time interval while the static password means the password is the same all the time when logging on to the system. The organization could force some requirements for the users to change the passwords periodically, e.g. weekly, monthly, or every half-year. The length of the time interval can be based on the sensitivity of the protected information. Zhang et al. [41] propose an algorithmic framework and empirical analysis of the security of modern password expiration.

*9) Use of Tokens*

The password can be generated using some security tokens that are used to ease authentication by authorized users of computer services. The token may be a physical device such as a smart card (it may also refer to a software token or virtual token). The password appearing on the token can be changed regularly with a time interval, which achieves the dynamic password mechanism and reduces the value of stolen passwords because of their short time validity. Furthermore, constantly shifting passwords reduce the probability of successful cracking by brute-force attack if the attacker uses the password list within a single shift. The challenge is how to deal with the synchronicity of the token and the server due to the fact that there will be a time delay before the password reaches the token from the server. After the user types in the password appearing on the token the password on the server site may be already changed to the next one due to the time delay. The token can be also equipped with inserted switch algorithms. In this case, the user types in the numbers appearing on the token then the calculator will generate a password using the inserted algorithms. Gyorffy et al. [42] propose a method of token-based graphical password authentication.

*10) Computer-Generated Passwords*

Users can also use computer-generated passwords for their accounts. Using some pre-design, the computer-generated password usually ensures a certain length, contains special characters, and is un-pronounceable, which is difficult for hackers to crack successfully within a short time. However, the computer-generated password is not easy to remember for the users due to the fact that it is mostly meaningless. Molloy and Li [43] introduce a work about the attack on the GridCode one-time password. Chiasson et al. [44] introduce the research on user interface design to affect security.

### *E. After the Generation of Passwords*

*11) Reactive Password Checking*

In a reactive password-checking strategy, the system periodically runs its own password cracker to find guessable passwords. The system will cancel passwords that are guessed and notify the users. The disadvantages are that it consumes resources very much and hackers can also use this strategy to find weak passwords if they get the password file copy[5]. Weir et al. [45] introduce a research work about testing metrics for password creation policies by attacking large sets of revealed passwords.

*12) Proactive Password Checker*

Another way to reject the weak passwords is proactive password checking. Different from reactive password checking, proactive password checking allows users to select their own passwords [46]. However, the system will check if the password is allowable or not. The goal is that users can select memorable passwords that are difficult to guess. Many researchers have designed their proactive password checkers. For example, Klein [47] proposes a proactive password checker to prevent easily guessed passwords from getting on the system in the first place according to the password cracking methods at that time, which includes trying the user’s relevant personal information (name, initials, account name, etc.), trying the words from various dictionaries, and the various permutations or capitalization permutations of the dictionary words, trying the foreign language words on foreign users and trying word pairs. Bishop and Klein [48] discuss a proactive tool for password checking called "passwd+". Bishop and Klein [8] design another proactive password checker “pwcheck” offering facilities to detect whether the passwords are easily guessed or not. The proposed password checker embraces some characteristics including rejecting any password in a set of common passwords (e.g. the dictionary words, passwords based on users’ name or account name and passwords that are shorter than a specific length), allowing per-user and per-site discrimination in its tests, having a pattern matching facility that can be stored in tests, being easy to set up, etc. Bergadano et al. [49] introduce high dictionary compression for proactive password checking. Ur et al. [50] make a discussion on the effect of strength meters that have been deployed by many websites to provide visual feedback on password strength and find that the stringent meters lead users to include more digits, symbols, and uppercase letters to strengthen the passwords.

*13) Password Encryption*

The password encryption protections include the hash functions and the use of salts. In computer security and cryptography, hash functions refer to the algorithms that take a variable-size input and return a fixed-size string output as the hash value. This approach ensures that any changes in the input data will result in a different hash value. There are several common characteristics of the hash function including easy computation (easy to compute the hash value for any given message), pre-image resistance (computationally infeasible to generate a message that has a given hash), second pre-image resistance (given an input A, it is difficult to find another different input B such that they have the same hash value), and collision resistance (computationally infeasible to find two different messages with the same hash), etc. There are some commonly used hash functions such as MD2, MD5[6] (produces a 128-bit hash value), and SHA (Secure Hash Algorithm specified in the Secure Hash Standard, which is developed by NIST and published as a federal information processing standard-FIPS PUB 180[7]). The SHA (0-3) algorithms are designed for the characteristics of fast computation and efficient implementation in hardware but these characteristics also make some disadvantages such as the ineffectiveness when preventing password cracking, such as the rainbow table attacks. Except for the encryption of the passwords, the hash function is also used mainly in digital signatures.

[5] http://www.cs.jhu.edu/~yairamir/cs418/os9/tsld021.htm

A "salt" in cryptography and computer security specifies random data (like the nonce) which is employed as the additional input to a one-way function that hashes a password. A new salt is generated randomly by the system for each new password. In general, a "salt" and a password are put together and encrypted with a hash function, and then the output hash value is stored in a file. The use of salt could efficiently slow down the pre-computed rainbow table attacks because the hackers have to hash each potential password traversing all the salts. The salts are widely used in different computer systems from the UNIX system credentials to internet security. For instance, Manber [51] designs two salts (one public and one secret) to protect passwords. The public salt is the same function as the salt at that time while the secret salt is generated randomly when the password is entered the first time and will be discarded by the designed system after use. The users do not need to know anything about the secret salt and the secret salt is not kept anywhere. Their experiments are tested on password schemes based on one-way functions. The performances show that using the designed two salts without changing any other part of the encryption mechanism can make password cracking more difficult 100 to 1000 times.

In the 1970s, the DES algorithm is used to encrypt passwords in the UNIX [52] with the later implications of DES encryption algorithm developed in the late 1980s [53] [54]. After that, the Unix systems replaced the traditional DES-based password hashing function "crypt()" with stronger methods such as "bcrypt" and "scrypt" [55]. These methods are also employed by other systems. For instance, the Cisco IOS uses MD5-crypt with 24-bit salt to encrypt password[8]. Large salt values are usually employed in these methods to reduce the efficiency of offline attacks due to the drastically increased cracking time. The MD5 and SHA-1 hash functions have been broken in recent years[9]. To achieve a long-term application of the hash functions, SHA-3 was designed and adopted as the NIST FIPS hash function standard in 2012[10].

*14) Access Control*

The general thought is that if the hackers cannot get the password (or encrypted password) files, then the password cracking efficiency will be highly reduced because the hackers cannot perform offline guessing. According to this analysis, password file access control is very important and efficient for the restraint of password cracking and this method is used in some systems. For instance, the modern UNIX operating systems store the hashed password file in the route that is accessible only to programs running with enhanced privileges [56]. Password file access control method also embraces some vulnerabilities, such as the weakness in the operating systems that allow access to the password file, accidental permission of making the password file readable, sniffing password transmission in the network, and weak challenge or response schemes in network protocols, etc. Steiner et al. [57] introduce work to secure a password-based cipher suite for transport layer security. Zhao and Yue [58] give an analysis of the security vulnerabilities in browser-based password managers, and propose a cloud-based storage-free design to achieve a high level of security with the desired confidentiality, integrity, and availability properties.

## VI. Conclusions and Future Work

The interaction of authentication, privacy, and integrity underlies the crucial aspects of the trust model that should be applied to be secured, of which the authentication for both users and data packets has been a well-known challenge [59]. Password remains the most widely used method to ensure authentication. However, there is some vulnerability in the passwords, such as offline attack, specific account attack, popular password attack, user analysis, workstation hijacking, user mistake exploitation, and electronic monitoring. Hackers usually utilize these vulnerabilities of the passwords to perform their attacks. This work first makes discussion of the widely used password-cracking methods and classifies them into several categories including the traditional methods (stealing, defrauding, user analysis, algorithm analysis, and fully guessing) and the combination with import technologies from other literature (Markov model and Probabilistic Context-Free Grammar). Then the countermeasures for password cracking are presented in two stages including the password design stage (e.g. user education, dynamic password, use of tokens and computer-generated password) and the protections after the generation of passwords (e.g. reactive password checking, proactive password checker, password encryption and password file access control). There are also some other countermeasures for password attacks such as intrusion detection measures, rapid re-issuance of compromised passwords, account lockout mechanisms, automatic workstation logout, and policies against similar passwords on network devices, etc.

In future work, we will introduce more advanced techniques on password cracking and countermeasures using computational language models developed recently from the natural language processing (NLP) and deep learning (DL) fields, such as RNNs and Transformers [65-67]. Some notable work can be found at [60-64].

## Acknowledgment

[6] http://tools.ietf.org/html/rfc1321
[7] http://www.itl.nist.gov/fipspubs/fip180-1.htm
[8] http://c3rb3r.openwall.net/mdcrack/download/FAQ-18.txt
[9] http://www.win.tue.nl/hashclash/rogue-ca/
[10] http://csrc.nist.gov/groups/ST/hash/sha-3/index.html

The author Han thanks Dr. Yain Whar SI, Lawrence from the University of Macau for his Computer Security courses.

## II. Appendix

### A. Appendix-A: Activity

To select a new, standard password hashing algorithm, a competition track is announced in 2013:

https://password-hashing.net/call.html

### B. Appendix-B: Some Available Software

*15) Cain and Abel password recovery tool:*

This is a tool for Microsoft Windows, which can recover many kinds of passwords with the employing of network packet sniffing, dictionary attacks, brute-force attacks, cryptanalysis attacks, and rainbow tables. [http://www.oxid.it/cain.html]

*16) John the Ripper password cracker:*

John the Ripper is free and Open Source software that is distributed primarily in source code. It is initially developed for Unix OS but now can run on different platforms, e.g. DOS, BeOS, Win32, etc. It combines different password crackers (including dictionary attack and brute-force modes) together with additional modules that can further extend its ability. [http://www.openwall.com/john/]

*17) Dave Grohl password cracker:*

It is a brute-force password cracker supporting dictionary attacks and distributed mode, which is suitable for all standard Mac OS X users. [http://www.davegrohl.org/]

*18) ElcomSoft Wireless Security Auditor:*

ElcomSoft Wireless Security Auditor under the company of ElcomSoft is a program that tries to guess the password associated with a wireless network using brute-force cracking methods and WPA/WPA2 Hash codes. [http://www.elcomsoft.com/]